\documentclass[aps,twocolumn,superscriptaddress]{revtex4-1}
\usepackage[colorlinks=true,bookmarks=true,citecolor=magenta,urlcolor=magenta,linkcolor=magenta,breaklinks]{hyperref} 
\usepackage{breakurl}
\usepackage{sidecap}
\usepackage{amssymb}
\usepackage{hhline}
\usepackage{urwchancal}
\sidecaptionvpos{figure}{t}
\usepackage{amsmath}
\usepackage{graphicx}
\usepackage{esint}
\usepackage{epstopdf}
\usepackage{rotating}
\graphicspath{{pict/}{./}}
\usepackage{bm}%
\usepackage{txfonts}
\usepackage[classicReIm]{kpfonts}

\newcounter{Fig}

\begin{document}


\title{Scattering Activities Bounded by Reciprocity and Parity Conservation}

\author{Weijin Chen}
\affiliation{School of Optical and Electronic Information, Huazhong University of Science and Technology, Wuhan, Hubei 430074, P. R. China}
\author{Qingdong Yang}
\affiliation{School of Optical and Electronic Information, Huazhong University of Science and Technology, Wuhan, Hubei 430074, P. R. China}
\author{Yuntian Chen}
\email{yuntian@hust.edu.cn}
\affiliation{School of Optical and Electronic Information, Huazhong University of Science and Technology, Wuhan, Hubei 430074, P. R. China}
\author{Wei Liu}
\email{wei.liu.pku@gmail.com}
\affiliation{College for Advanced Interdisciplinary Studies, National University of Defense
Technology, Changsha, Hunan 410073, P. R. China}

\begin{abstract}
Scattering activities are generally manifest through different optical responses of scattering bodies to circularly polarized light of opposite handedness. Similar to the ubiquitous roles played by scattering theory across different branches of photonics, scattering activities can serve as a fundamental concept to clarify underlying mechanisms of various chiroptical effects, both within and beyond scattering systems. In this work we investigate scattering activities for reciprocal systems that exhibit various geometric symmetries but are intrinsically achiral. We reveal how scattering activities are generally bounded by reciprocity and parity conservation,  demonstrating that though extinction activity is usually eliminated by symmetry, scattering activities in forms of distinct absorptions, scatterings or angular scattering patterns can more widely emerge. Since our analyses are solely based on fundamental laws of reciprocity and parity conservation, regardless of geometric and optical parameters of the scattering systems studied, the principles revealed are generically applicable. The intuitive and pictorial framework we have established is beyond any specific coupling models, able to reveal hidden connections between seemingly unrelated chiral manifestations, and thus more accessible for a unified understanding of various chiroptical effects.
\end{abstract}

\maketitle

\section{Introduction}
Chiroptical effects in general correspond to different responses of optical systems to circularly polarized (CP) waves of opposite handedness, that is, left and right handed circularly  polarized (LCP and RCP) light~\cite{BARRON_2009__Molecular,BEROVA_2011__Comprehensive,COLLINS_AdvancedOpticalMaterials_chirality_2017,CALOZ_2019_ArXiv190309087Phys._Electromagnetic}.
Historically, optical activity represents a special kind of chiroptical effect that is synonymous with rotations of polarization planes of linearly polarized light during propagation. In a more general modern sense, optical activities and chiroptical effects can be used on equal terms, both of which can be traced to the same origin of different responses to incident LCP and RCP waves~\cite{BARRON_2009__Molecular,BEROVA_2011__Comprehensive,COLLINS_AdvancedOpticalMaterials_chirality_2017,CALOZ_2019_ArXiv190309087Phys._Electromagnetic}. Optical activities can be roughly divided into two categories of intrinsic and extrinsic ones:  for intrinsic activities the optical systems themselves are chiral which cannot be superimposed on their mirror-imaging counterparts; while for extrinsic activities, the optical systems are achiral, while the whole systems can be viewed as chiral when the incident waves are also considered~\cite{WILLIAMS_1969_J.Chem.Phys._Optical,PLUM_2009_J.Opt.A:PureAppl.Opt._Extrinsic}. Intrinsic optical activities have already been extensively studied over more than one hundred years, with its fundamentals well clarified~\cite{BARRON_2009__Molecular,BEROVA_2011__Comprehensive}. In contrast, extrinsic optical activities had almost remained in oblivion until the recent emergence and rapid expansion of metamaterials and its related disciplines, thanks to which enormous attention has been attracted in the past few years~\cite{COLLINS_AdvancedOpticalMaterials_chirality_2017,PLUM_2009_J.Opt.A:PureAppl.Opt._Extrinsic,LI_2013_J.Opt._Chiral,OH_2015_NanoConvergence_Chiral,HENTSCHEL_2017_Sci.Adv._Chiral,
PAPAKOSTAS_Phys.Rev.Lett._optical_2003,KUWATA-GONOKAMI_2005_Phys.Rev.Lett._Giant,FEDOTOV_Phys.Rev.Lett._asymmetric_2006,SCHWANECKE_2008_NanoLett._Nanostructured,
DREZET_2008_Opt.ExpressOE_Optical,PLUM_Phys.Rev.Lett._metamaterials_2009,EFTEKHARI_2012_Phys.Rev.B_Strong,SCHAFERLING_2012_Phys.Rev.X_Tailoring,
HOPKINS_2016_LaserPhotonicsRev._Circular,KHANIKAEV_NatCommun_experimental_2016,HU_2017_Sci.Rep._Alldielectrica,YE_2017_Phys.Rev.Applied_Large,NAJAFABADI_2017_ACSPhotonics_Analytical,NAJAFABADI_2017_SciRep_Optical,HU_2017_Sci.Rep._Alldielectric,ZHU_2018_LightSciAppl_Giant,
WANG_2018_LightSciAppl_Reflective,MA_2018_Opt.ExpressOE_Alldielectric,DAVIS_2019_Sci.Adv._Microscopic,XIE__Adv.Opt.Mater._Lattice}. Though various simple models have been constructed to explain extrinsic optical activities (or hybrid activities of both sorts) in specific structures~\cite{PLUM_2009_J.Opt.A:PureAppl.Opt._Extrinsic,PLUM_Phys.Rev.Lett._metamaterials_2009,SCHAFERLING_2012_Phys.Rev.X_Tailoring,DAVIS_2019_Sci.Adv._Microscopic,NAJAFABADI_2017_ACSPhotonics_Analytical,
KHANIKAEV_NatCommun_experimental_2016,EFTEKHARI_2012_Phys.Rev.B_Strong}, confusions and misconceptions still exist and are widely scattered among existing literatures (refer to the recent work Ref.~\cite{DAVIS_2019_Sci.Adv._Microscopic} for updated discussions about the situation). A unified framework beyond specific models and substantiated by general principles are badly needed for clarifications, which unfortunately is still unavailable.

In optics and photonics, Mie scattering serves as a fundamental concept since most investigations can be ultimately traced back to scattering related problems~\cite{Bohren1983_book}.  As an outstanding example, it is known that the responses of periodic structures (in both regimes with and without extra diffractions beyond transmissions and reflections) are dictated by the scattering properties of the unit-cell constituents~\cite{LIU_2018_Opt.Express_Generalized,LIU_Phys.Rev.Lett._generalized_2017,CHEN_2019__Singularities}: transmissions decided by forward scatterings; reflections and other diffractions decided by scatterings at corresponding diffraction angles; and non-radiative losses decided by absorption cross sections.  Scattering activities can be manifest in the forms of different extinction, scattering and absorption cross sections, or distinct angular scattering patterns with respect to incident LCP and RCP waves. It is expected that comprehensive investigations into scattering activities would serve as an essential step for a unified framework for coherent understandings of all chiroptical effects.

\begin{figure*}[tp]
\centerline{\includegraphics[width=14cm]{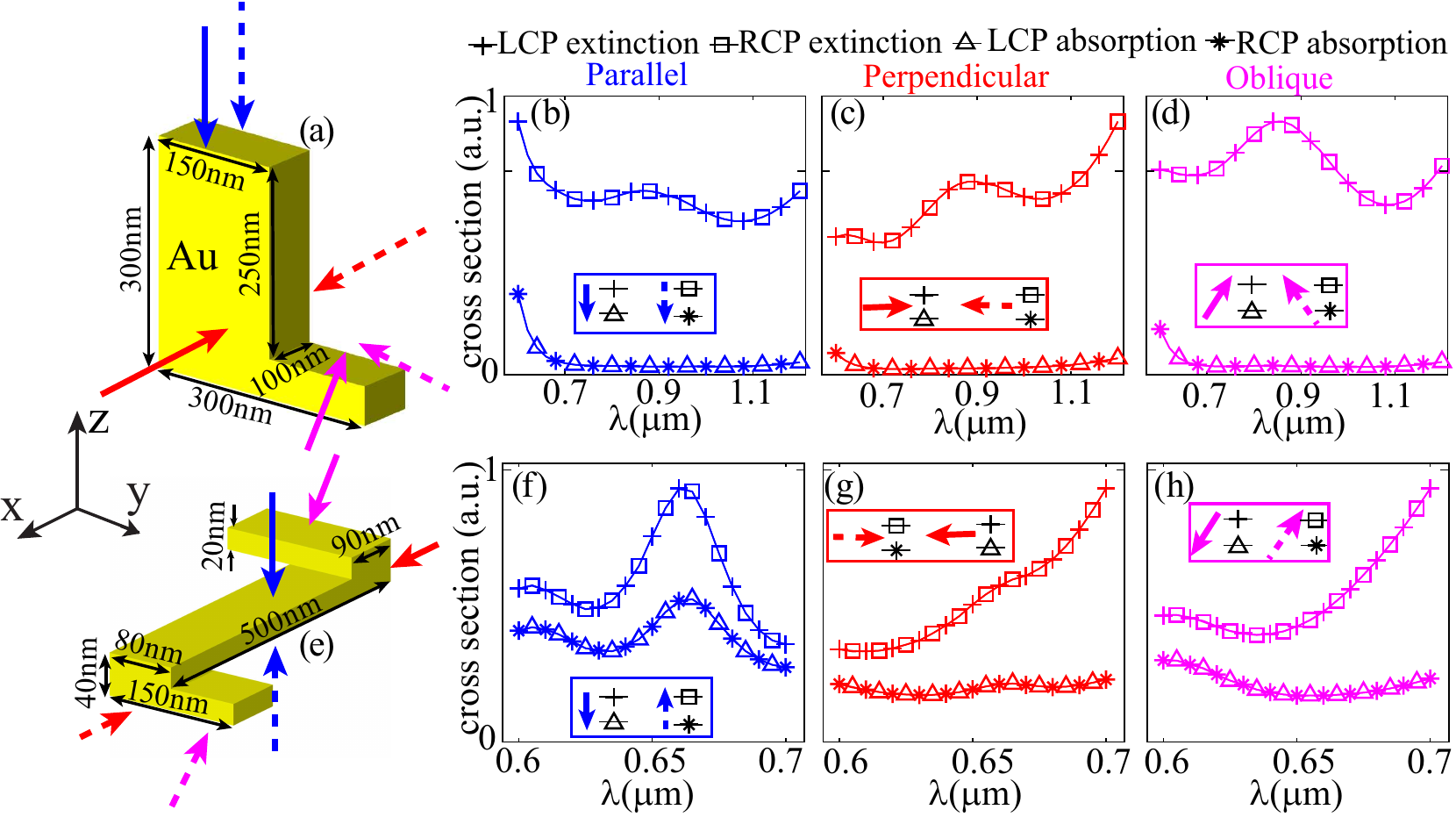}} \caption {\small Two scattering configurations involving gold particles (with geometric parameters directly specified) that exhibit mirror symmetry in (a) or inversion symmetry in (e), where the incident waves are indicated by solid arrows (LCP) or dashed arrows (RCP). The arrows in the same color represent CP waves that are either mirror-imaging (a) or point-imaging (e) counterparts. The extinction and absorption spectra are summarized in (b)-(d) for the mirror symmetry case in (a), and in (f)-(h) for the inversion symmetry case in (e). The three sets of spectra for each symmetry scenario correspond to three pairs of parity-imaging processes incident from different directions (with respect to the $\textbf{y-z}$ plane) in (a) and (e): parallel (blue), perpendicular (red) and oblique (purple). The wavevector $\textbf{k}$ makes an angle of $45^{\circ}$ with respect to the horizontal $\textbf{x-y}$ plane ($\textbf{k}\perp \textbf{y}$) for the oblique incidence.}\label{fig1}
\end{figure*} 

Here we study extrinsic scattering activities of achiral scattering systems that exhibit different geometric symmetries but no intrinsic chirality. Solely based on fundamental laws of reciprocity and parity conservation, we demonstrate that extinction activities are widely eliminated by reciprocity and parity conservation, while activities in terms of different absorption and scattering cross sections can generally survive. Moreover, even if all cross sections are identical for LCP and RCP waves, activities can still show in the form of different angular scattering patterns. Our framework is beyond specific mode-coupling models, formulas-free and fully pictorial, revealing general principles that are independent of specific geometric or optical parameters of the scattering bodies studied. As will be shown, it can further reveal hidden connections between seemingly unrelated  chiroptical manifestations. Our work can provide coherently simple while penetrating perspectives for broader communities to obtain a unified understanding of chiroptical effects, in both the linear and nonlinear optical regimes~\cite{COLLINS_AdvancedOpticalMaterials_chirality_2017,COLLINS_2019_Phys.Rev.X_First}.

\section{Scattering Invariance Dictated by Parity Conservation and Reciprocity}

\subsection{Invariance of Extinction, Scattering and Absorption Induced by Parity Conservation}

Throughout our work the investigations are conducted in a three dimensional (3D) setting as all realistic scattering structures are rigourously 3D no matter how thin they are and how they can be effectively viewed as two dimensional (2D). In 3D, valid parity operations include mirror operations (flip the sign of one spatial coordinate) and inversion (point-imaging) operations (flip the signs of all three spatial coordinates)~\cite{BARRON_2009__Molecular,LEE_1956_Phys.Rev._Question}.  For CP light, any parity operation would flip the handedness of the wave and change the propagation direction (represented by vector \textbf{k}) accordingly: an inversion operation would flip it to its opposite \textbf{-k} and a mirror operation would change it to its mirror-imaging vector. The law of parity conservation tells that electromagnetic interactions can not possibly differentiate the original optical process from its point- or mirror-imaging (parity-paired) processes~\cite{BARRON_2009__Molecular,LEE_1956_Phys.Rev._Question}. It means that all cross sections of extinction, scattering and absorption would remain the same after parity operations and scattering patterns would be transformed to their parity-paired equivalents.  This law is universally applicable as long as electroweak interactions are not involved, and thus fully independent of specific geometric or optical parameters~\cite{BARRON_2009__Molecular,LEE_1956_Phys.Rev._Question}.

\begin{figure*}[tp]
\centerline{\includegraphics[width=15cm]{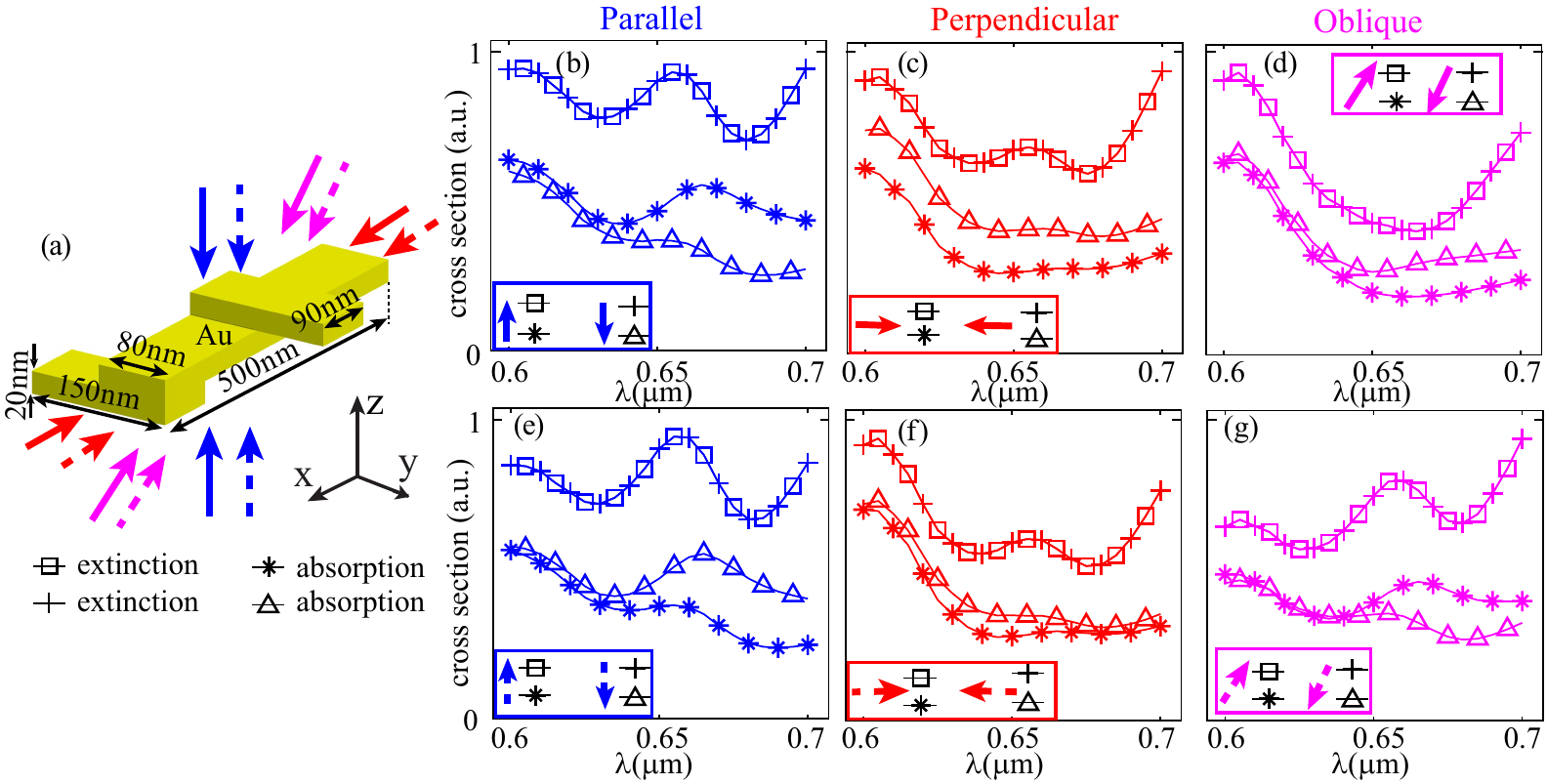}} \caption {\small (a) The scattering of CP waves by a gold particle that exhibit neither mirror nor inversion symmetry. Waves incident from opposite directions with the same handedness constitute a reciprocity-pair for which the extinctions are identical guaranteed by reciprocity. In (a) six such reciprocity-pairs are shown: one LCP pair (solid arrows) and one RCP pair (dashed arrows) for each scenario: parallel (blue), perpendicular (red) and oblique (purple) incidences (with respect to the vertical $\textbf{y-z}$ plane). The extinction and absorption spectra for each pair are summarized in (b)-(g). The wavevector $\textbf{k}$ makes an angle of $45^{\circ}$ with respect to the horizontal $\textbf{x-y}$ plane ($\textbf{k}\perp \textbf{y}$) for the oblique incidence.}\label{fig2}
\end{figure*} 

Direct manifestations of parity conservation in optical scattering processes of CP waves are showcased in Fig.~\ref{fig1}.  The scattering particle shown in Fig.~\ref{fig1}(a) exhibits vertical-mirror symmetry, and this mirror operation will not change the particle but would flip the handedness of the incident waves (solid arrows and dashed arrows represent LCP and RCP waves, respectively) and mirror-image \textbf{k} (indicated by directions of the arrows) accordingly (see a pair of dashed and solid arrows in the same color). Here the handedness is defined according to the spatial (temporal) ration trend viewed along (opposite to) \textbf{k}, and we employ the $\exp(\omega t-\textbf{k}\textbf{r})$ notation for the electromagnet waves.  The propagation direction is maintained (flipped) when light is incident along (perpendicular) to the mirror symmetry plane (\textbf{y-z} plane), and obliquely incident waves will be transformed to the other side of the same incident angle with respect to the \textbf{y-z} plane.  For specific calculations and demonstrations, we assume that the particle is made of gold (with permittivity taken from Ref.~\cite{Johnson1972_PRB}) and the geometric parameters are specified in the figure (as is the case throughout this work), though the general laws and principles to be revealed are independent of those specific parameters. For all possible incident angles, parity conservation requires that the extinction, absorption and scattering cross sections are identical for a pair of mirror-imaging processes, as is shown in Figs.~\ref{fig1}(b)-(d) for three different incidence scenarios.  Here we show only extinction and absorption cross sections (obtained through numerical calculations using COMSOL Multiphysics), and the scattering cross sections can be directly obtained through subtracting absorptions from extinctions according to the optical theorem~\cite{Bohren1983_book}.

The spectra for a gold scattering particle exhibiting inversion symmetry shown in Fig.~\ref{fig1}(e) are summarized in Fig.~\ref{fig1}(f)-(h), where for a pair of point-imaging processes all the crosse sections are identical, as is also required by parity conservation.  Similar to the aforementioned mirror symmetry scenario,  an inversion operation would flip the handedness of incident wave. But the difference is that, regardless of the incident angles, an inversion operation would always flip \textbf{k} to \textbf{-k}. We note here that, though in Fig.~\ref{fig1} we investigate only particles that are invariant under parity operations, parity conservation does not require the scattering particles to exhibit any symmetry. That is to say, for arbitrary scattering particles, a parity operation can be always implemented, with identical observable effects for parity-paired physical processes. The problem is that, when there is no required symmetry, the scattering particle would not be invariant after parity operations, ending up with a distinct scattering system irrelevant to our following discussions of scattering activities.

\subsection{Extinction Invariance Induced by Reciprocity}

As a next step, we proceed to show how reciprocity is manifested during scattering processes. This principle is applicable to reciprocal scattering bodies: their permittivity and permeability tensors are symmetric; or equivalently there are no magneto-optical effects, nonlinearity or temporal modulations~\cite{FAN_science_comment_2012,CALOZ_2018_Phys.Rev.Applied_Electromagnetica}. A recent study confirms that reciprocity is directly incarnated into extinction invariance for scattering bodies of arbitrary shapes, as long as the incident waves are of the same handedness and opposite \textbf{k}~\cite{SOUNAS_Opt.Lett.OL_extinction_2014}. The application of reciprocity, and of parity conservation with an inversion operation (or a perpendicular-mirror operation: mirror operation with \textbf{k} perpendicular to the mirror) have two common features: opposite \textbf{k} of incident waves and identical extinction. The differences are also obvious: reciprocity requires the incident waves of the same handedness and put no constraints on absorption and scattering cross sections; while parity conservation with inversion or perpendicular-mirror operations requires CP waves of opposite handedness and guarantees invariance of all cross sections. In Fig.~\ref{fig2}(a) we show a scattering configuration with a gold particle without any mirror or inversion symmetries. The above claims are further demonstrated in Figs.~\ref{fig2}(b)-(g), where the effects of extinction invariance secured by reciprocity are clear for each pair of incident CP waves that are connected by reciprocity (opposite \textbf{k} and same handedness). The extinction and absorption spectra for six such pairs (two pair for each incidence scenario, which could be parrel, perpendicular or oblique to the vertical $\textbf{y-z}$ plane) are summarized in Figs.~\ref{fig2}(b)-(g). It is also shown that the absorption and thus also scattering cross sections are not bounded by reciprocity, which could be different for each pair of incidences.

\section{Scattering Activities Bounded by Reciprocity and Parity Conservation}

\subsection{Absence of Scattering Activities in All Forms Induced by Mirror Symmetry}

After the demonstrations of reciprocity and parity conservation during light scattering processes, we proceed to study how scattering activities are restricted by those general laws. Strictly speaking, the emergence of scattering activities have two necessary preconditions: besides opposite handedness, the incident CP waves should be of the same propagation direction. A scattering system responding differently to LCP and RCP waves of different propagation directions are irrelevant to scattering activities. In this respect, all the previous analyses about parity conservation and reciprocity are not directly related to scattering activities. The only exception is the scenario with \textbf{k} parallel to the reflection mirror when the parity operation [parallel-mirror operation; see Fig.~\ref{fig1}(a)] maintains \textbf{k}.  Basically, parity conservation with parallel-mirror operation requires all responses to be identical to LCP and RCP light, eliminating scattering activities of any forms [refer to Fig.~\ref{fig1}(b)].

\subsection{Absence of Extinction Activities Induced by Reciprocity and Parity Conservation}

\begin{figure}[tp]
\centerline{\includegraphics[width=8.5cm]{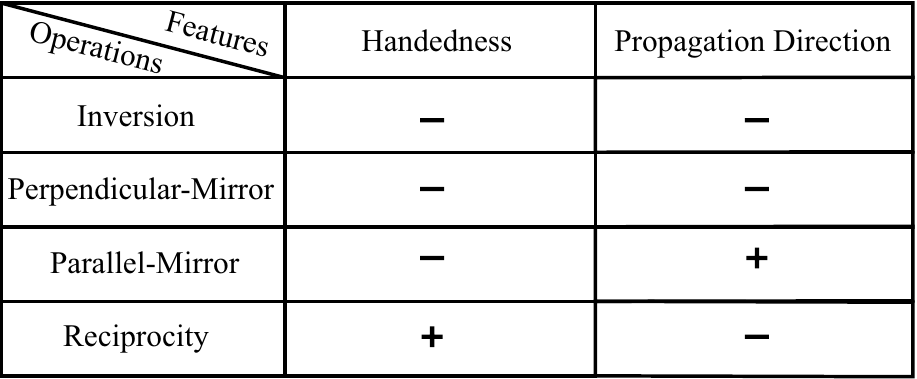}} \caption{\small The effects of different operations on propagation directions (or wavevector \textbf{k}) and handedness of incident CP waves. This includes the reciprocity operation and three parity operations: inversion, parallel-mirror and perpendicular-mirror operations. The signs ``$\textbf{+}$" and ``$\textbf{-}$" indicate maintaining and flipping of \textbf{k} or handedness, respectively.}\label{fig3}
\end{figure}

Now we seek to establish a connection between scattering activities and our previous analyses concerning reciprocity and parity conservation. For convenience of discussions, we define the reciprocity operation which flips \textbf{k} but maintain the handedness of incident waves, and leaves the scattering bodies unchanged.  The reciprocity operation is the same as the time-reversal operation with regard to incident CP waves, but for the scattering bodies they can be totally different since time-reversal operation would transform the losses to gain. In this language of operations, reciprocity guarantees the extinction invariance under reciprocity operation but put no constraints on responses in terms of scattering and absorption cross sections. The results with regard to how different operations affect \textbf{k} or handedness of incident waves are summarized in Fig.~\ref{fig3}, where ``$\textbf{+}$" and ``$\textbf{-}$" indicate maintaining and flipping of \textbf{k} or handedness, respectively. Moreover, all parity operations leave the extinction, scattering and absorption unchanged, while the reciprocity operation requires only the invariance of extinction.

According to Fig.~\ref{fig3}, the combination of a proper parity operation (inversion or perpendicular-mirror operation) and a reciprocity operation would flip the handedness while maintain \textbf{k} of incident CP waves, as is required for activity analysis. Since both operations secure extinction invariance, extinction activity would not emerge, provided that the scattering bodies are operation-invariant (the combined operations would transform the scattering body back to itself). This requirement of operation-invariance is as natural as that of LCP and RCP to be of the same \textbf{k}, since different responses of different scattering configurations also have nothing to do with activities. Considering that a reciprocity operation does not change the scattering body, for reciprocal scattering systems of either mirror symmetry (\textbf{k} is required to be perpendicular to the symmetry mirror) or inversion symmetry (for arbitrary \textbf{k}), extinction activity would be absent. At the same time, for both scenarios activities of scattering and absorption are not bounded as reciprocity operations are involved.

\begin{figure}[tp]
\centerline{\includegraphics[width=7.5cm]{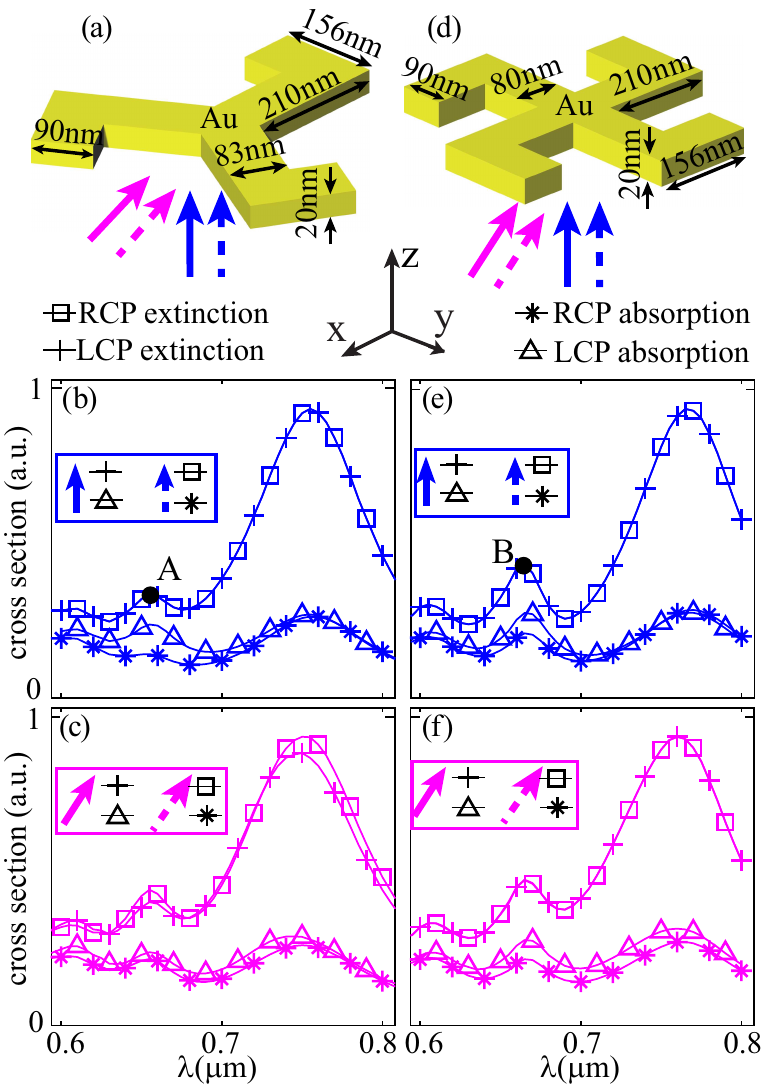}} \caption{\small Two scattering configurations involving gold particles that exhibit mirror symmetry in (a) or both mirror and inversion symmetry (d), where the incident waves are indicated by solid arrows (LCP) or dashed arrows (RCP). The extinction and absorption spectra are summarized in (b)-(c) for the the configuration in (a), and in (e)-(f) for that shown in (d). Two sets of spectra for each scenario correspond to two different incident directions: perpendicular incidence along $\textbf{z}$ and oblique incidence, with $\textbf{k}$ making an angle of $45^{\circ}$ with respect to the horizontal $\textbf{x-y}$ plane ($\textbf{k}\perp \textbf{y}$). Two points are marked  ($\lambda=0.66~\mu$m) in (b) and (e) for further investigations shown in Figs.~\ref{fig7}(d) and (e).}
\label{fig4}
\end{figure}

To confirm what has been claimed above, we start with a scattering gold particle shown in Fig.~\ref{fig4}(a), which exhibits mirror symmetry (with respect to the horizontal \textbf{x-y} plane). The extinction and absorption spectra are summarized in  Figs.~\ref{fig4}(b) and (c), for perpendicularly and obliquely incident CP waves, respectively. As expected, the extinction activity is absent for perpendicular incidence while for oblique incidence it would emerge, since for the latter case a combination of reciprocal and  horizontal mirror operations does not bring \textbf{k} to itself. For both cases, activities in terms of absorption (and thus scattering) are not bounded and clearly visible.

Then we turn to another gold scattering particle shown in Fig.~\ref{fig4}(d) which exhibits also inversion symmetry. For such a configuration, regardless of the incident directions, the extinction activities are always absent, with no constraints on activities of absorptions and scatterings [refer to the spectra shown in Figs.~\ref{fig4}(e) and (f) for two different incident angles]. Relying on the framework we establish here upon general laws of reciprocity and parity conservation, it now becomes rather simple to comprehend the related results obtained in previous studies~\cite{KUWATA-GONOKAMI_2005_Phys.Rev.Lett._Giant,EFTEKHARI_2012_Phys.Rev.B_Strong,HOPKINS_2016_LaserPhotonicsRev._Circular,KHANIKAEV_NatCommun_experimental_2016,NAJAFABADI_2017_ACSPhotonics_Analytical,NAJAFABADI_2017_SciRep_Optical,ZHU_2018_LightSciAppl_Giant,DAVIS_2019_Sci.Adv._Microscopic}, especially those results related to planar activity that requires simultaneous oblique incidence and in plane inversion symmetry breaking~\cite{PLUM_2009_J.Opt.A:PureAppl.Opt._Extrinsic,PAPAKOSTAS_Phys.Rev.Lett._optical_2003,PLUM_Phys.Rev.Lett._metamaterials_2009}. For example in Ref.~\cite{HOPKINS_2016_LaserPhotonicsRev._Circular}, though the scenarios of oblique incidences are not investigated, the inversion symmetry of the structures studied there would guarantee the absence of extinction activities for arbitrarily incident angles of CP waves.

\subsection{Absence of Extinction Activities for Scattering Bodies with Improper Rotation Symmetry}

\begin{figure}[tp]
\centerline{\includegraphics[width=7.5cm]{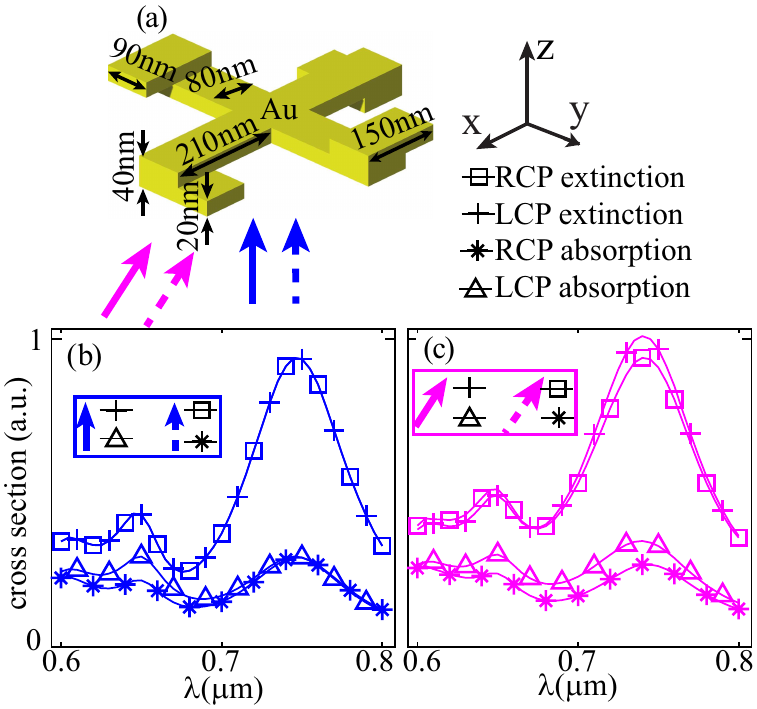}} \caption{\small (a) The scattering of LCP (solid arrows) and RCP (dashed arrows) waves by a gold particle that exhibit $S_4$ improper rotation symmetry. The extinction and absorption spectra are summarized in (b)-(c) for perpendicular (along $\textbf{z}$) and oblique ($\textbf{k}$ makes an angle of $45^{\circ}$ with respect to the horizontal $\textbf{x-y}$ plane with $\textbf{k}\perp \textbf{y}$) incidences, respectively.}
\label{fig5}
\end{figure}

A special feature of a CP wave is that it remains invariant after a ration operation (by any angle) along \textbf{k}, with neither its propagation direction nor the handedness changed. That is to say, rotate the scattering bodies along \textbf{k} will change none of the cross sections of extinction, scattering or absorption. As a result, add an extra operation after the aforementioned combined reciprocal and perpendicular-mirror operations will maintain the extinction invariance.  It is now clear that, if a scattering body remains invariant after a improper rotation operation (combined operations of a mirror operation and a ration operation along an axis perpendicular to the mirror) and light is incident along the rotation axis, extinction activity would also be eliminated, without the requirement of mirror or inversion symmetry.  A specific scattering configuration is provided in Fig.~\ref{fig5}(a), where the gold scattering particle exhibit $S_4$ symmetry (remains invariant after a four-fold improper rotation operation). The spectra summarized in Figs.~\ref{fig5}(b) and (c) indicate that there is no extinction activity for perpendicular incidence along \textbf{z} while for another oblique incidence, the extinction activity would arise. For both scenarios, activities in terms of absorption (and thus scattering) are not restricted and clearly visible, similar to the scenarios shown in Fig.~\ref{fig4}. The conclusions we draw here are also applicable for scattering bodies with improper rotation symmetry of higher orders $S_{2n}$ where $n>2$.

\subsection{Scattering Activities Manifested Through Different Scattering Patterns}

\begin{figure}[tp]
\centerline{\includegraphics[width=9cm]{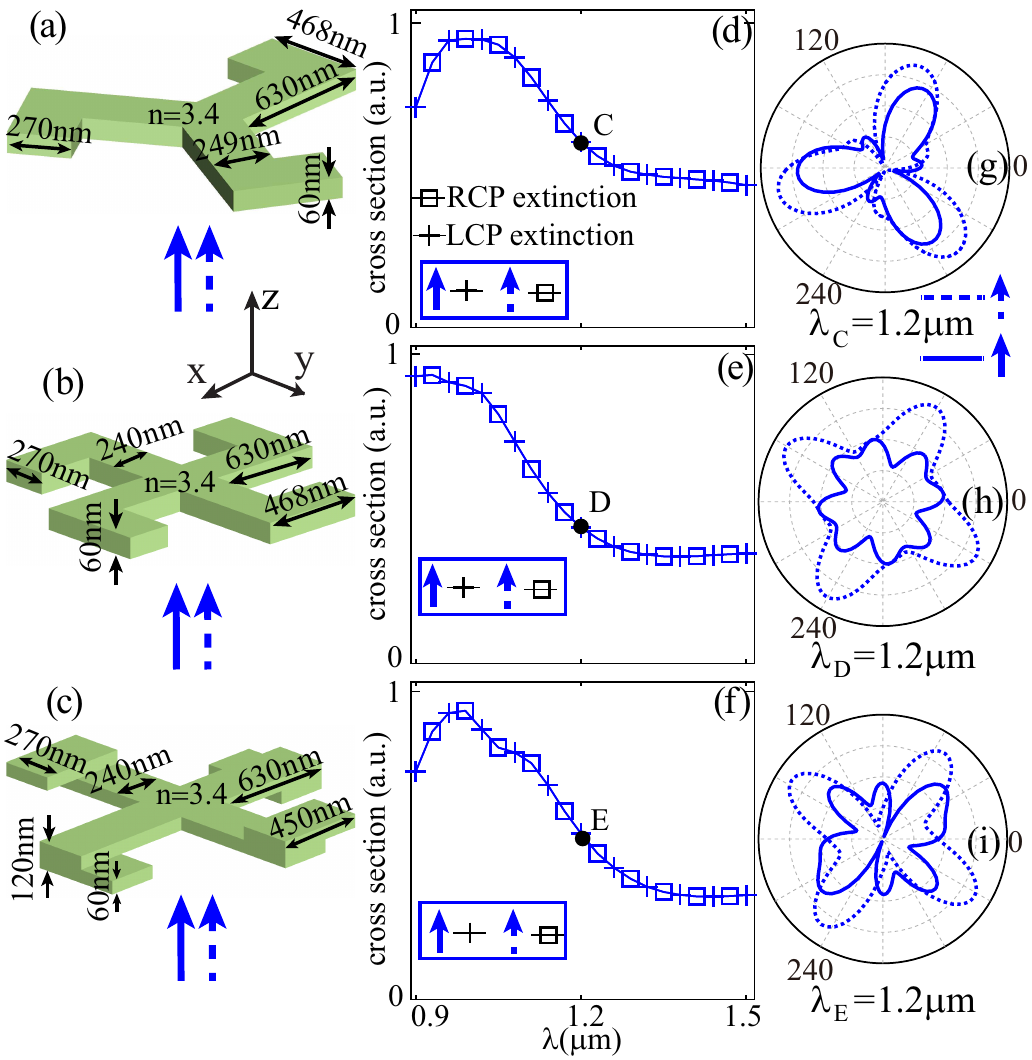}} \caption{\small Three scattering particles that exhibit the same symmetries as those investigated in Figs.~\ref{fig4} and \ref{fig5} are shown in (a)-(c), with the material of lossy gold changed to lossless dielectric material of refractive index $3.4$. The LCP (solid arrows) and RCP (dashed arrows) waves are incident along $\textbf{z}$ and the extinction spectra are shown in (d)-(f) respectively. For each case one point is indicated ($\lambda=1.2~\mu$m), for which the angular scattering patterns on the $\textbf{x-y}$ plane for incident waves of opposite handedness (solid lines: LCP; dashed lines: RCP) are shown correspondingly in (g)-(i).}
\label{fig6}
\end{figure}

Up to now, we have discussed only lossy scattering bodies. What would happen if they are losses and thus absorption cross sections are zero? According to optical theorem~\cite{Bohren1983_book}, the absence of extinction activity would mean the elimination of all possible activities in terms of cross sections. But still the lossless scattering bodies can respond differently to LCP and RCP waves through distinct angular scattering patterns. In Figs.~\ref{fig6}(a)-(c) we show scattering configurations that possess identical  symmetries to those studied in Figs.~\ref{fig4} and \ref{fig5} without any extinction activities, except that now the lossy gold is changed to losses dielectric material of refractive index $3.4$. For those all-dielectric scattering bodies, though activities in terms of extinction and scattering cross sections (absorptions are zero; extinctions and scattering cross sections are equal) are absent [see Figs.~\ref{fig6}(d)-(f)], activities in terms of scattering patterns emerge, as are summarized in Fig.~\ref{fig6}(g)-(i) showing angular scattering patterns on the $\textbf{x-y}$ plane, at the corresponding marked positions in Figs.~\ref{fig6}(d)-(f) of wavelength $\lambda=0.7~\mu$m.

\begin{figure}[tp]
\centerline{\includegraphics[width=7cm]{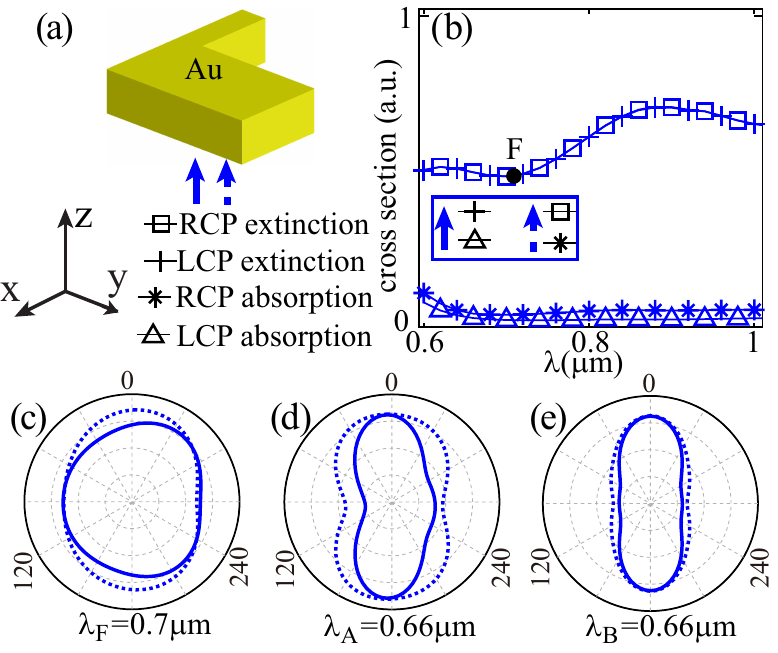}} \caption{\small (a) The scattering of LCP (solid arrows) and RCP (dashed arrows) waves by a gold particle [identical to that investigated in Fig.~\ref{fig1}] that exhibits only mirror symmetry, with neither inversion nor rotation (either proper or improper rotation) symmetry. The extinction and absorption spectra are summarized in (b) for the perpendicular (along $\textbf{z}$) incidence. The angular scattering patterns on the $\textbf{x-z}$ plane at the marked point ($\lambda=0.7~\mu$m) of (b) and other two points marked in Figs.~\ref{fig4}(b) and (e) are shown in (c)-(e). For each case, the forward scattering direction corresponds to zero polar angle.}
\label{fig7}
\end{figure}

The absence of extinction activities demonstrated in Fig.~\ref{fig4} with perpendicular incidences are guaranteed by mirror symmetries only, with nothing to do with the obvious rotation symmetries exhibited by the scattering bodies. In Fig.~\ref{fig7}(a) we show a much simpler configuration of a gold scattering particle which possesses only mirror symmetry, and as expected there is no extinction activity [Fig.~\ref{fig7}(b)] for perpendicular incidence along \textbf{z} direction. Similar structures have been extensively studied in previous studies and the same conclusions have been drawn~\cite{EFTEKHARI_2012_Phys.Rev.B_Strong,HOPKINS_2016_LaserPhotonicsRev._Circular,KHANIKAEV_NatCommun_experimental_2016,NAJAFABADI_2017_ACSPhotonics_Analytical,NAJAFABADI_2017_SciRep_Optical,DAVIS_2019_Sci.Adv._Microscopic}, though based on more complicated arguments compared to what we have presented in this work.  Unfortunately, there is a widely spread misconception that extinction invariance directly means forward-scattering invariance (see discussions in Ref.~\cite{NAJAFABADI_2017_ACSPhotonics_Analytical} for example), since extinction is induced by destructive interferences between the incident waves and the forward scatted waves~\cite{Bohren1983_book}.  A key point that has been usually neglected is that interferences do not occur between waves of orthogonal polarizations, and thus identical extinction means only the same amount of forward scattered waves that are of the same polarization as the incident wave. That is to say, extinction has nothing to do with orthogonal forward scatterings. For incident CP waves specifically, extinction does not provide any information about forward scattering waves of opposite handedness as they do not contribute to the extinction.

For the scattering body shown in Fig.~\ref{fig7}(a), there is linear birefringence and thus part of the incident CP waves can be converted to orthogonal waves of opposite handedness along the forward direction~\cite{YARIV_1984__Optical,DECKER_2010_Opt.Lett._Twisted}. Consequently, though at the marked point  ($\lambda=0.7~\mu$m) of the spectra in Fig.~\ref{fig7}(b) the extinction is identical for RCP and LCP waves, the overall forward scatterings are not the same [shown in Fig.~\ref{fig7}(c) at the polar angle of zero]. Another related misconception is that the same extinction is always accompanied by the same transmission. This is wrong due to the fact that though orthogonal forward scatterings do not contribute to extinction, they do contribute to the overall transmission. Based on those arguments, the previous demonstrations of handedness-dependent transmissions (or a seemingly different but essentially equivalent version of asymmetric transmissions for oppositely propagating CP waves of the same handedness) without any extinction activities become more comprehensible~\cite{FEDOTOV_Phys.Rev.Lett._asymmetric_2006,SCHWANECKE_2008_NanoLett._Nanostructured,HU_2017_Sci.Rep._Alldielectrica,YE_2017_Phys.Rev.Applied_Large,MA_2018_Opt.ExpressOE_Alldielectric}, which are induced by linear birefringence that results in circular cross-polarization conversions.

In contrast, the n-fold ($n\geq 3$) rotational symmetry along the incident axis of the configurations studied in Fig.~\ref{fig4} (with perpendicular incidences) fully suppresses the linear birefringence and thus prevent circular cross-polarization conversions~\cite{DECKER_2010_Opt.Lett._Twisted,FERNANDEZ-CORBATON_2013_Opt.ExpressOE_Forwarda}. For those scenarios, all forward scattered waves are of the same polarization as the incident waves and thus all contribute to the extinction. This means that the absence of extinction activities would result in identical forward scatterings for LCP and RCP incident waves, as has been confirmed by the scattering patterns shown in Figs.~\ref{fig7}(d) and (e), at the points marked in Figs.~\ref{fig4}(b) and (e)  ($\lambda=0.66~\mu$m). In a similar fashion, with n-fold ($n\geq 3$) rotational symmetry along the incident axis, the absence of extinction activity would also lead to identical transmissions of RCP and LCP waves (or equivalently, identical transmissions for oppositely propagating CP waves of the same handedness~\cite{COLLINS_AdvancedOpticalMaterials_chirality_2017,PLUM_2009_J.Opt.A:PureAppl.Opt._Extrinsic,FEDOTOV_Phys.Rev.Lett._asymmetric_2006}.

\section{Conclusions and Discussions}
To summarize, we discuss activities of achiral scattering bodies under the the general laws of reciprocity and parity conservation. Scattering activities can be manifested through not only cross sections of extinction, scattering and absorptions, but also different angular scattering patterns. It is demonstrated that, though the combination of reciprocity and parity conservation widely eliminates extinction activities, activities in the forms of different cross sections of absorption and scattering can generally survive. For lossless scattering bodies, even if all cross sections are identical for LCP and RCP waves, the activity can still emerge through different scattering patterns. Our pictorial and more accessible analysis are fully based on general laws and thus free free specific mode coupling models and their associated complex formulas. As a result, the principles we have revealed are generic and widely applicable, which constitute a simple and coherent framework for establishing connections between various chiroptical effects. This unified framework we have established can potentially bring new perspectives for further studies into activities, especially when other exotic effects are incorporated, such as PT symmetry, nonlinearity, and non-hermitian effects~\cite{YU_2019_Curr.Opt.Photon.COPP_Chiralitya,DROULIAS_2019_Phys.Rev.Lett._Chiral}.

We acknowledge the financial support from National Natural Science Foundation of China (Grant No. 11874026 and 11874426), and the Outstanding Young Researcher Scheme of National University of Defense Technology.



\end{document}